\documentclass[aps,%
final,%
notitlepage,%
oneside,%
twocolumn,
nobibnotes,%
nofootinbib,%
superscriptaddress,%
showpacs,%
centertags,
floatfix]%
{revtex4}
\usepackage[dvips]{graphicx}
\usepackage{epsfig,graphics,graphicx}
\usepackage{amsmath,amsthm,amssymb,amscd}

\begin{document}

\title{Entanglement of quantum circular states of light}

\author{D.~B.~Horoshko}
\affiliation{Univ. Lille, CNRS, UMR 8523 - PhLAM - Physique des Lasers Atomes et Mol\'ecules, F-59000 Lille, France}
\affiliation{B.~I.~Stepanov Institute of Physics, NASB, Nezavisimosti Ave.~68, Minsk 220072 Belarus}%

\author{S.~De~Bi\`evre}
\affiliation{Univ. Lille, CNRS, UMR 8524 - Laboratoire Paul Painlev\'e, F-59000 Lille, France}
\affiliation{Equipe-Projet MEPHYSTO,Centre de Recherche INRIA Futurs,
Parc Scientifique de la Haute Borne, 40, avenue Halley B.P. 70478,
F-59658 Villeneuve d'Ascq cedex, France.}

\author{M.~I.~Kolobov}
\affiliation{Univ. Lille, CNRS, UMR 8523 - PhLAM - Physique des Lasers Atomes et Mol\'ecules, F-59000 Lille, France}

\author{G.~Patera}
\affiliation{Univ. Lille, CNRS, UMR 8523 - PhLAM - Physique des Lasers Atomes et Mol\'ecules, F-59000 Lille, France}

\date{\today}

\begin{abstract}
We present a general approach to calculating the entanglement of formation for superpositions of two-mode coherent states, placed equidistantly on a circle in the phase space. We show that in the particular case of rotationally-invariant circular states the Schmidt decomposition of two modes, and therefore the value of their entanglement, are given by analytical expressions. We analyse the dependence of the entanglement on the radius of the circle and number of components in the superposition. We also show that the set of rotationally-invariant circular states creates an orthonormal basis in the state space of the harmonic oscillator, and this basis is advantageous for representation of other circular states of light.
\end{abstract}

\maketitle

\section{Introduction}
Coherent states of light play a central role in quantum optics, being representatives of deterministic classical waves in the quantum formalism, and serving as an effective basis for various quasiprobability approaches to the description of the quantum field \cite{ScullyZubairy97}. Quantum superpositions of two coherent states have been intensively investigated in the last three decades, especially for the case of distinct macroscopic amplitudes of two  states, where the superposition can be considered as an optical analog of the famous ``Schr\"odinger cat'' state of a quantum system \cite{Dodonov74,Yurke86,Buzek92,Horoshko97}. Experimental realizations of such superpositions reach mesoscopic values of the amplitude both in microwave \cite{Raimond97,Kirchmair15} and optical \cite{Takahashi08,Ourjoumtsev09} domains and allow one to study the elusive border between the quantum and the classical worlds.

\begin{figure}[ht]
 \includegraphics[width=0.5\columnwidth]{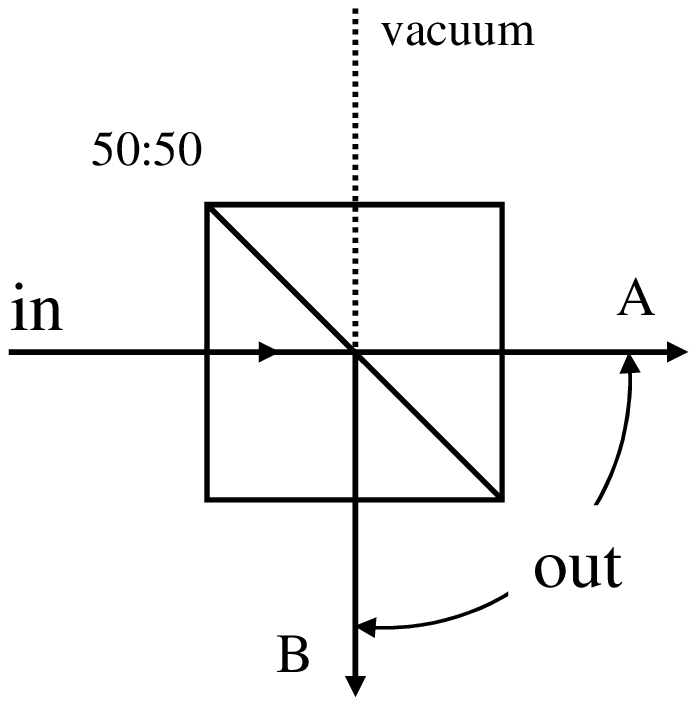}
\caption{\label{BS} A symmetric beam splitter. The ``in-state'' of the mode A is transformed into an (in general) entangled ``out-state'' of the modes A and B.}%
\end{figure}

Another interesting feature of a single-mode coherent superposition is its simple transformation into a two-mode coherent superposition by means of a beam splitter (see Fig.\ref{BS}):
\begin{equation}\label{trans}
\sum_{m=0}^{N-1}c_m|\sqrt{2}\alpha_m\rangle_A|0\rangle_B \rightarrow \sum_{m=0}^{N-1}c_m|\alpha_m\rangle_A|\alpha_m\rangle_B,
\end{equation}
where $\left\{|\sqrt{2}\alpha_m\rangle_A\right\}$ are coherent states with complex amplitudes $\{\sqrt{2}\alpha_m\}$ at one input to the beam-splitter (mode A), $\{c_m\}$ are arbitrary complex numbers satisfying the normalisation condition, $|0\rangle_B$ is a vacuum state at another input to the beam-splitter (mode B), and the coherent states at both outputs of a 50:50 beam-splitter have amplitudes $\alpha_m$, which are all different. When the number of components is higher than 1, the output state is an entangled state of two modes, a coherent state superposition \cite{Sanders92}. Such states are of considerable interest and great practical importance for quantum communication and computation \cite{vanEnk03,Kilin11} (for a review, see Ref.\cite{Sanders12}). In particular one is interested in quantifying their entanglement by calculating the entanglement of formation $E=-Tr\{\rho_A\log_2\rho_A\}$, where $\rho_A$ is the partial density operator of the mode A alone \cite{Bennett96}.

In the case where the individual coherent states $|\alpha_m\rangle$ are macroscopically distinguishable, i.e. $|\alpha_m-\alpha_n|\gg 1$, $m\ne n$, the right-hand side of Eq.(\ref{trans}) represents almost a Schmidt decomposition \cite{Law00} with the entanglement given by the Shannon entropy of the coefficients $|c_m|^2$, reaching $\log_2 N$ for an equally-weighted superposition. In the case where not all of the coherent states are well distinguishable, the expression for the entanglement of formation is known only for the case of two components, since for this case the partial density operator is represented in a proper basis by a $2\times2$ matrix, and its eigenvalues can be readily found in an analytic form by solving a second-order characteristic equation.

In the present article we deduce an analytic expression for the entanglement for the case of arbitrary number $N$ of components in the superposition in a particular case of coherent states $|\alpha_m\rangle$ placed equidistantly on a circle of radius $|\alpha_0|$ centered at the origin, and having a linear relative phase dependence. For this particular case of ``rotationally-invariant circular states'' (RICS) we find also the explicit form for the Schmidt basis, given by a set of other RICS.

The article is structured as follows. In Section II we discuss the class of circular states of light for a single mode of field and concentrate on the subclass of single-mode RICS. In Section III we develop a general formalism for calculating the entanglement of a superposition of two-mode coherent states, apply it to the case of two-mode circular states, and show that in the case of a two-mode RICS the Schmidt decomposition is obtained in a simple analytic form. In Section IV we show that a set of RICS forms a basis in the space of circular states having advantages for representation of any circular state of light. In Section V we discuss the questions of entanglement optimization and quantification of non-classicality on the basis of the obtained results.

\section{Single-mode states}
\subsection{Circular states}
\begin{figure}[ht]
\centerline{\includegraphics[width=1.0\columnwidth]{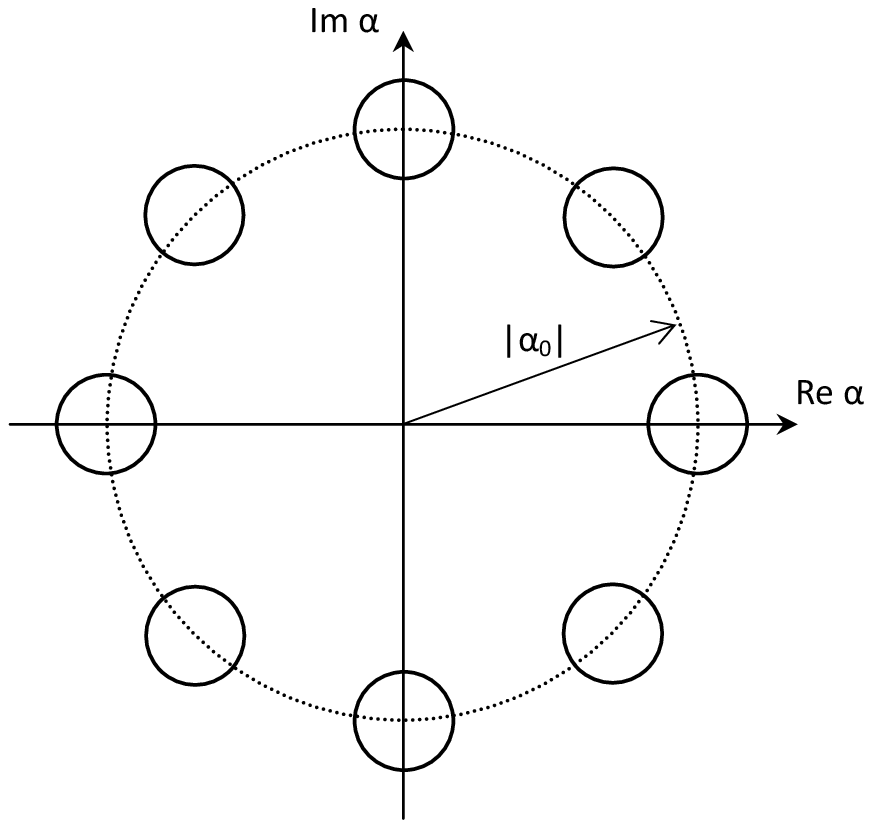}}
\caption{\label{Fig0} Coherent states on the circle of radius $|\alpha_0|$. Each coherent state is represented by a circle of radius $\frac12$, corresponding to the $\sigma$-area of its Wigner function, being a two-dimensional Gaussian distribution.}%
\end{figure}

We consider one mode of the electromagnetic field, represented by a quantum harmonic oscillator. The quantum state of our interest is a superposition of $N$ coherent states $\{|\alpha_m\rangle, m=0,1,...,N-1\}$, placed equidistantly on the circle of radius $|\alpha_0|$ (see Fig.\ref{Fig0}):
\begin{equation}\label{circ}
|\psi_c\rangle=\sum_{m=0}^{N-1}c_m\vert\alpha_m\rangle,
\end{equation}
where
\begin{equation}\label{alpham}
\alpha_m=\alpha_0 e^{-i2\pi m/N},
\end{equation}
$\alpha_0$ is an arbitrary non-zero complex number, and $\{c_m\}$ are arbitrary complex coefficients, satisfying the normalization condition $\langle\psi_c|\psi_c\rangle=1$.

The states of this class are often called ``circular states'' \cite{Jose00,Wang10} and were extensively studied during the last decades in connection with their highly non-classical properties \cite{Dodonov74,Yurke86,Buzek92,Horoshko97,Raimond97,Kirchmair15,Takahashi08,Ourjoumtsev09}. They are also favorite test benches for studying decoherence in elementary quantum systems both theoretically \cite{Zurek01,Horoshko98} and experimentally \cite{Raimond97}. An important subclass is represented by generalized coherent states of light possessing a certain phase-periodicity property \cite{Birulia68}, including a family of the so-called ``Kerr states'', which can be produced unitarily in a third-order nonlinear optical process of self-phase modulation \cite{Yurke86,Kirchmair15,Tanas91,Tara93}. Another important subclass is discussed in the next subsection.

All circular states are eigenstates of the $N$th power of the photon annihilation operator:
\begin{equation}\label{power}
a^N|\psi_c\rangle=\alpha_0^N|\psi_c\rangle.
\end{equation}

The set of coherent states with amplitudes given by Eq.(\ref{alpham}), is characterized by its Gram matrix $G_{mn}=\langle \alpha_m|\alpha_n\rangle$, which has the following cyclic form
\begin{equation}\label{GN}
G_{mn}=g(m-n),
\end{equation}
where
\begin{equation}\label{g}
g(m)=\exp\left\{|\alpha_0|^2\left(e^{i2\pi m/N}-1\right)\right\}
\end{equation}
is an $N$-component vector of inner products, and here and below all lowercase discrete variables are taken modulo $N$, if not otherwise specified. We note that $g^*(m)=g(-m)$, as required by the Hermiticity of the Gram matrix.

As we will see later, an important role in the analysis of these states is played by discrete Fourier transform in the $N$-dimensional complex linear space, which is defined as
\begin{equation}\label{DFT}
\tilde g(k)=\frac1{N}\sum_{m=0}^{N-1}g(m)e^{-i2\pi km/N},
\end{equation}
with the inverse transform
\begin{equation}\label{DFTinv}
g(m)=\sum_{k=0}^{N-1}\tilde g(k)e^{i2\pi km/N}.
\end{equation}
Below we will always denote the Fourier transform of any function $\{f(m),m=0,1,...,N-1\}$ by the same letter with a tilde: $\tilde f(k)$, and imply that its argument takes values $0\le k\le N-1$.

Decomposing the exponent in Eq.(\ref{g}) into Taylor series and taking the Fourier transform Eq.(\ref{DFT}), we obtain
\begin{equation}\label{tildeg}
\tilde g(k)=e^{-|\alpha_0|^2}\sum_{l=0}^\infty \frac{|\alpha_0|^{2(k+lN)}}{(k+lN)!},
\end{equation}
i.e. $\tilde g(k)$ is given by a sum of weights of values $j$ from a Poisson distribution, such that $(j\mod N) = k$. As a consequence, all $\tilde g(k)$ are positive and sum to unity.

\subsection{Rotationally-invariant circular states}

Each state of the set  $\{|\alpha_m\rangle, m=0,1,...,N-1\}$ with amplitudes given by Eq.(\ref{alpham}), is produced from the previous one in this set by a rotation in phase space  $|\alpha_m\rangle=U_N|\alpha_{m-1}\rangle$, described by the unitary operator
\begin{equation}\label{U}
\hat U_N = e^{-i2\pi a^\dagger a/N},
\end{equation}
where $a$ is the photon annihilation operator. Thus, the $N$-dimensional space of circular states, determined by Eqs.(\ref{circ},\ref{alpham}) is invariant under the action of this operator. Let us find the eigenstates of $\hat U_N$ in this space. The eigenvalues of a unitary operator have the form $e^{-i\varphi}$, where $\varphi$ is real. Substituting Eq.(\ref{circ}) into $U_N|\psi_{c}\rangle=e^{-i\varphi}|\psi_{c}\rangle$ and taking into account the linear independence of the states $\{|\alpha_m\rangle, m=0,1,...,N-1\}$ when $\alpha_0\ne0$, we obtain  $c_{m+1}=e^{i\varphi} c_{m}$. Starting with $m=0$ by iteration we find $c_m=c_0e^{im\varphi}$. After the $N$th iteration we arrive at the condition of periodicity $\exp(iN\varphi)=1$, meaning that $\varphi$ should be such that $N\varphi=2\pi q$, where $q$ is an integer. Thus, the eigenvalues of $U_N$ are given by $e^{-i2\pi q/N}$. There are only $N$ different eigenvalues, for which we let $0\le q \le N-1$. The corresponding eigenstates are

\begin{equation}\label{alphaNm}
|\mathrm{c}_q\rangle =  \frac1{N\sqrt{\tilde g(q)}}\sum_{m=0}^{N-1}e^{i2\pi mq/N}|\alpha_0 e^{-i2\pi m/N}\rangle.
\end{equation}
where the normalization factor is written explicitly with $\tilde g(q)$ defined by Eq.(\ref{DFT}). The eigenstates satisfy the orthonormality condition
\begin{equation}\label{orthog}
\langle \mathrm{c}_q|\mathrm{c}_r\rangle = \delta_{qr},
\end{equation}
and are invariant (up to a global phase) under rotation with the operator $U_N$:
\begin{equation}\label{eigen}
U_N|\mathrm{c}_q\rangle = e^{-i2\pi q/N}|\mathrm{c}_q\rangle.
\end{equation}

The last property allows us to call them rotationally-invariant circular sates (RICS), while Eq.(\ref{orthog}) shows that these states form an ortonormal basis in the space of circular states, defined by Eq.(\ref{circ}).

The states Eq.(\ref{alphaNm}) create an important subclass in the class of circular states, Eq.(\ref{circ}). For $N=2$ they give ``even'' ($q=0$) and ``odd'' ($q=1$) coherent states of single-mode field \cite{Dodonov74}, extensively investigated in the last decades \cite{Buzek92,Horoshko97,Raimond97,Takahashi08,Ourjoumtsev09}. The ``compass state'', a model for studying the decoherence of a one-dimensional quantum system \cite{Zurek01}, is also a member of this class for $N=4$, $q=0$. In general form the states Eq.(\ref{alphaNm}) were studied by Janszky and co-workers \cite{Janszky93,Janszky95}, who suggested a method of their generation by means of cavity QED \cite{Domokos94}.

Using the decomposition of a coherent state in the Fock basis, we rewrite the RICS, Eq.(\ref{alphaNm}), in the form \cite{Janszky95}
\begin{equation}\label{cq}
|\mathrm{c}_q\rangle =  \frac{e^{-|\alpha_0|^2/2}}{N\sqrt{\tilde g(q)}}\sum_{l=0}^{\infty}\frac{\alpha_0^{q+lN}}{\sqrt{(q+lN)!}}|q+lN\rangle,
\end{equation}
where $|q+lN\rangle$ is a Fock state with $q+lN$ photons. Eq.(\ref{cq}) shows that a RICS is a sum of Fock states with the number $j$ of photons such, that $(j\mod N)=q$. In the case of $N=2$ this property is reduced to a fixed parity of the photon number, peculiar to ``even'' and ``odd'' coherent states.

When $N\gg|\alpha_0|^2$, one can leave only the first terms in the sums Eqs.(\ref{tildeg},\ref{cq}). In this way one obtains a Fock state with $q$ photons for infinite number of components or for the amplitude tending to zero:
\begin{equation}\label{lim}
\lim_{N\rightarrow \infty}|\mathrm{c}_q\rangle = \lim_{\alpha_0\rightarrow 0}|\mathrm{c}_q\rangle = |q\rangle.
\end{equation}

\section{Two-mode states}
\subsection{Superpositions of two-mode coherent states}

In this section we come back to the problem of entanglement of two-mode states formulated in the Introduction. We consider two modes of optical field A and B represented by quantum harmonic oscillators. An arbitrary two-mode coherent state has the form $|\alpha\rangle_A|\beta\rangle_B$ where $|\alpha\rangle_A$ and $|\beta\rangle_B$ are coherent states for modes A and B respectively with some complex amplitudes $\alpha$ and $\beta$. We consider only the symmetric case $\beta=\alpha$, a generalization to a more general case being straightforward. An arbitrary superposition of $N$ symmetric two-mode coherent states can be written as:
\begin{eqnarray}\label{Psid}
|\Psi_c\rangle_{AB}&=&\sum_{m=0}^{N-1}c_m|\alpha_m\rangle_A|\alpha_m\rangle_B\\\nonumber
&=&\left(\begin{array}{ccc}|\alpha_0\rangle_A &...& |\alpha_{N-1}\rangle_A\end{array}\right) \hat C \left(\begin{array}{c}|\alpha_0\rangle_B\\...\\|\alpha_{N-1}\rangle_B\end{array}\right),
\end{eqnarray}
where $\hat C$ is a complex diagonal $N\times N$ matrix of coefficients, $C_{mn}=c_m\delta_{mn}$, and the coherent amplitudes for the discussion of this and next subsections can be arbitrary (all different), though for the case of our interest they are given by Eq.(\ref{alpham}).

It has been pointed out in the Introduction, that splitting a single mode superposition on a beam-splitter results in a two-mode superposition of coherent states, Eq.(\ref{trans}). Now we can read this equation ``from right to left'' and state that any (symmetric) two-mode superposition of coherent states, Eq.(\ref{Psid}), can be considered as being produced by splitting the corresponding single-mode state
\begin{eqnarray}\label{in}
|\psi_c^{in}\rangle_{A}&=&\sum_{m=0}^{N-1}c_m|\sqrt{2}\alpha_m\rangle_A,
\end{eqnarray}
on a 50:50 beamsplitter with vacuum at the other input. It is important to note that the coefficients of the two-mode coherent states at the output are the same as that of the single-mode coherent states at the input, and that the normalization to unity is reached for the states Eq.(\ref{Psid}) and Eq.(\ref{in}) for the same set of coefficients $\{c_m\}$. Below we will refer to the state Eq.(\ref{in}) as the``in-state'' of the two-mode superposition, Eq.(\ref{Psid}). In practice, a superposition state can be created by various means, not necessarily by beam-splitting, but the concept of the corresponding ``in-state'' proves to be highly useful for understanding the properties of entanglement of such superpositions, as is shown below. In this connection we will use the same generic name for a two-mode superposition as for its single-mode ``in-state'', i.e. we will speak about a ``two-mode circular state'' or a ``two-mode RICS'' if their corresponding ``in-states'' are given by Eq.(\ref{circ}) or Eq.(\ref{alphaNm}) respectively. It is easy to see that a two-mode RICS is invariant under simultaneous rotation of two modes with the operator $e^{-i2\pi a^\dagger a/N} e^{-i2\pi b^\dagger b/N}$, where $a$ and $b$ are photon annihilation operators for modes A and B respectively, justifying thus the extension of the name.

The entanglement of the two-mode state, Eq.(\ref{Psid}), is expected to be determined by the Gram matrix of the set of states $\hat G$ and by the matrix of coefficients $\hat C$. When the set contains only one component ($N=1$), entanglement is zero. The corresponding ``in-state'', Eq.(\ref{in}), is in general highly nonclassical and its non-classicality is characterized by the Gram matrix of the set of states, which is a scaled version of $\hat G$, and by its matrix of coefficients, which coincides with $\hat C$. When the set contains only one component ($N=1$), the ``in-state'' is classical (coherent). We see that non-classicality of the single-mode ``in-state''  Eq.(\ref{in}) is closely related to the entanglement of the two-mode ``out-state'' Eq.(\ref{Psid}). This analogy can be used for the quantification of non-classicality in the spirit of the well-known approach of ``entanglement potential'' \cite{Asboth05}, and will be discussed below in Section V.

\subsection{Orthogonalization in the general case}
The set of coherent states $\{|\alpha_m\rangle,m=0,...,N-1\}$ is non-orthogonal and spans some $N$-dimensional subspace $\mathcal{S}_N$. We may look for expressing it via states of an orthonormal basis $\{|v_0\rangle,...,|v_{N-1}\rangle\}$ in $\mathcal{S}_N$:
\begin{eqnarray}\label{Basis}
\left(\begin{array}{ccc}|\alpha_0\rangle &...& |\alpha_{N-1}\rangle\end{array}\right)&=& \left(\begin{array}{ccc}|v_0\rangle &...& |v_{N-1}\rangle\end{array}\right)\hat L,
\end{eqnarray}
where $\hat L$ is a complex $N\times N$ matrix. This matrix is non-unitary in general and satisfies
\begin{eqnarray}\label{Gram}
\hat L^{\dagger}\hat L = \left(\begin{array}{c}\langle \alpha_0|\\...\\\langle \alpha_{N-1}|\end{array}\right)
\left(\begin{array}{ccc}|\alpha_0\rangle &...& |\alpha_{N-1}\rangle\end{array}\right)\equiv \hat G,
\end{eqnarray}
where $\hat G$ is the Gram matrix for the set of coherent states.

In the orthonormal basis (for both modes) Eq.(\ref{Psid}) can be rewritten as
\begin{eqnarray}\label{Psid2}
|\Psi\rangle_{AB}&=&\left(\begin{array}{ccc}|v_0\rangle_A &...& |v_{N-1}\rangle_A\end{array}\right) \hat L \hat C \hat L^T\left(\begin{array}{c}|v_0\rangle_B\\...\\|v_{N-1}\rangle_B\end{array}\right),
\end{eqnarray}
and the partial density operator of the mode A reads as
\begin{eqnarray}\label{rhod}
\rho_A&=&\left(\begin{array}{ccc}|v_0\rangle_A &...& |v_{N-1}\rangle_A\end{array}\right) \hat M \hat M^{\dagger} \left(\begin{array}{c}{}_A\langle v_0|\\...\\{}_A\langle v_{N-1}|\end{array}\right),
\end{eqnarray}
where $\hat M=\hat L \hat C \hat L^T$. Now the problem of determining the entanglement of the state Eq.(\ref{Psid}) is reduced to the problem of finding the eigenvalues of the matrix $\hat M \hat M^{\dagger}=\hat L\hat C\hat G^T\hat C^\dagger\hat L^\dagger$. These eigenvalues can always be found numerically for any finite $N$. However, an analytic solution of the eigenvalue problem represents an important theoretical problem, giving us the eigenbasis in addition to the eigenvalues of the partial density operator and providing a better understanding of the underlying physics.

The form of the matrix $\hat M \hat M^{\dagger}$, is determined by three issues: the structure of the set of coherent states, represented by the Gram matrix $\hat G$, the relative weights and phases, represented by the matrix $\hat C$, and the working basis, represented by the matrix $\hat L$. The first two matrices will have highly symmetric forms for the states of our interest. As for the last matrix, we will use the symmetrization procedure introduced by L\"owdin \cite{Lowdin50} for which the transform matrix is Hermitian and is given by  $\hat L = \hat L^{\dagger} = \hat G^{1/2}$. The orthonormal basis obtained in this way is characterized by minimal distance from the non-orthogonal set \cite{Mayer} and the whole orthogonalization procedure is symmetric with respect to the initial set. Application of this particular orthogonalization procedure is justified by a successful deduction of the Schmidt decomposition on its ground.

\subsection{Orthogonalization for two-mode circular states}

Now we consider the set of coherent states with amplitudes given by Eq.(\ref{alpham}), whose Gram matrix is given by Eq.(\ref{Gram}). The eigenvalue equation for the Gram matrix reads
\begin{equation}\label{eigeneq}
\sum_{j=0}^{N-1}\hat G_{ij}v_j =\lambda v_i,
\end{equation}
where $\{v_j,j=0,1,...,N-1\}$ is the eigenvector. In the Fourier domain this equation reads
\begin{equation}\label{eigeneqf}
\left(N\tilde g(k)-\lambda\right)\tilde v_k =0, \quad \forall k,
\end{equation}
where we have taken into account the cyclic structure of the Gram matrix.

Eq.(\ref{eigeneqf}) has $N$ solutions given by $\lambda^{(n)}=N\tilde g(n)$ with the corresponding eigenvectors $\tilde v_k^{(n)}=\delta_{kn}$, or in the direct domain
\begin{equation}\label{eigenv}
v_k^{(n)} = e^{i2\pi kn/N}.
\end{equation}
Now we build a unitary matrix $\hat V_{kn}=v_k^{(n)}$ having the eigenvectors as columns (it follows from Eq.(\ref{eigenv}) that this matrix is symmetric), and write
\begin{equation}\label{Gdecomp}
\hat G = \hat V\hat\Lambda\hat V^\dagger,
\end{equation}
where $\hat\Lambda$ is a diagonal matrix of eigenvalues, $\hat\Lambda_{mn}=N\tilde g(n)\delta_{mn}$. Now the matrix of transform to the L{\"o}wdin basis is $\hat L = \hat V\hat\Lambda^{1/2}\hat V^\dagger$ with the matrix elements
\begin{equation}\label{L}
\hat L_{mn} = \sum_{k=0}^{N-1}\sqrt{N\tilde g(k)}e^{i2\pi k(m-n)/N},
\end{equation}
and the superposition matrix $\hat M = \hat L\hat C\hat L^T$ has the elements
\begin{eqnarray}\label{MN}
\hat M_{mn}&=&\sum_{jk}\sqrt{\tilde g(j)}\tilde c_{j+k}\sqrt{\tilde g(k)}e^{i2\pi (jm+kn)/N},
\end{eqnarray}
where $\tilde c_k$ is the Fourier transform of the  vector of superposition coefficients $c_m$.

We notice that the superposition matrix can be written as $\hat M = \hat V\hat M'\hat V$, where $\hat M' = \hat\Lambda^{1/2}\hat V^*\hat C\hat V^*\hat\Lambda^{1/2}$, the asterisk standing for complex conjugation and  the symmetricity of the matrix $\hat V$ having been used. Now the partial density matrix of mode $A$ is $\hat R=\hat M\hat M^\dagger= \hat V\hat R'\hat V^\dagger$, where the matrix $\hat R'=\hat M'\hat M'^\dagger$ has the elements
\begin{equation}\label{Ra}
\hat R'_{mn} = N^2 \sqrt{\tilde g(m)\tilde g(n)} \sum_{k=0}^{N-1}\tilde c_{m+k}\tilde g(k)\tilde c^*_{k+n}.
\end{equation}
Since the matrices $R$ and $R'$ are related by a unitary transformation, they have the same eigenvalues. Thus, the problem of computing the entanglement of the superposition state is reduced to the problem of finding the entropy of the distribution given by the eigenvalues of the matrix $R'$.

Another object of interest are the eigenvectors of the partial density matrix (the Schmidt basis). They can be found from Eq.(\ref{rhod}) rewritten as
\begin{eqnarray}\label{rhod2}
\rho_A&=&\left(\begin{array}{ccc}|v'_0\rangle_A &...& |v'_{N-1}\rangle_A\end{array}\right) \hat R' \left(\begin{array}{c}{}_A\langle v'_0|\\...\\{}_A\langle v'_{N-1}|\end{array}\right),
\end{eqnarray}
where the basis $\{|v'_n\rangle\}$ is the Fourier transform of the L{\"o}wdin basis $\{|v_n\rangle\}$:
\begin{eqnarray}\label{Basis2}
\left(\begin{array}{ccc}|v'_0\rangle &...& |v'_{N-1}\rangle\end{array}\right)&=& \left(\begin{array}{ccc}|v_0\rangle &...& |v_{N-1}\rangle\end{array}\right)\hat V \\\nonumber &=&\left(\begin{array}{ccc}|\alpha_0\rangle &...& |\alpha_{N-1}\rangle\end{array}\right) \hat L^{-1}\hat V.
\end{eqnarray}

\subsection{Schmidt decomposition for two-mode RICS}

It is readily seen from Eq.(\ref{Ra}) that both the eigenvalues and the eigenvectors of the $\hat R'$ matrix can be found in analytic form in the case of $\tilde c_k=\mathcal{C}\delta_{kq}$, where $\mathcal{C}$ is a real number. This case corresponds to a two-mode RICS, whose ``in-state'' is given by Eq.(\ref{alphaNm}) with a replacement $\alpha_0\rightarrow\sqrt{2}\alpha_0$. Therefore, the normalization constant is $\mathcal{C}=(N\sqrt{\tilde g_1(q)})^{-1}$, where $\tilde g_1(q)$ is a Fourier transform of $g_1(m)=g^2(m)$, which are the elements of the ``scaled'' Gram matrix for the set of states $\{|\sqrt{2}\alpha_m\rangle, m=0,1,...,N-1\}$. Substituting $\tilde c_k=\mathcal{C}\delta_{kq}$ into Eq.(\ref{Ra}) we obtain a diagonal matrix $\hat R'$:
\begin{equation}\label{Ra2}
\hat R'_{mn} = \frac{\tilde g(n)\tilde g(q-n)}{\tilde g_1(q)} \delta_{mn},
\end{equation}
whose eigenvectors are given by Eq.(\ref{Basis2}). Taking into account that $\hat L^{-1}\hat V = \hat V \hat\Lambda^{-1/2}$, we obtain from Eqs.(\ref{alphaNm},\ref{Basis2})
\begin{equation}\label{vprime}
|v'_n\rangle = |\mathrm{c}_n\rangle,
\end{equation}
that is, the Schmidt basis is given by the Fourier-transformed L\"owdin basis, coinciding with the corresponding family of RICS. Application of the operator  $\hat V \hat\Lambda^{-1/2}$ to the set of coherent states means that we first take a Fourier transform of the set and then normalize it. If we took an inverse Fourier transform afterwards, we would obtain the true L{\"o}wdin basis, which is close to the initial set of coherent states for the case of their good separation.

Now we consider the RICS $|\mathrm{c}_q\rangle$ with parameters $\{N,\sqrt{2}\alpha_0\}$ at one input of a 50:50 beam-splitter with the vacuum at the other one. It follows from Eq.(\ref{alphaNm}) and Eq.(\ref{trans}) that the state of the two output modes is
\begin{equation}\label{out0}
|\Psi_c\rangle_{AB} = \frac{1}{N\sqrt{\tilde g_1(q)}}\sum_{m=0}^{N-1}e^{i2\pi qm/N}|\alpha_m\rangle_A|\alpha_m\rangle_B,
\end{equation}
where the amplitudes for both modes are given by Eq.(\ref{alpham}). On the other hand, it follows from Eqs.(\ref{Ra2},\ref{vprime}), considered in the framework of the previous subsection, that the state at the output of the beam-splitter can be written in its Schmidt form as
\begin{equation}\label{out2}
|\Psi_c\rangle_{AB} = \frac{1}{\sqrt{\tilde g_1(q)}}\sum_{k=0}^{N-1}\sqrt{\tilde g(k)\tilde g(q-k)}|\mathrm{c}_k\rangle_A|\mathrm{c}_{q-k}\rangle_B.
\end{equation}
The normalization of the state Eq.(\ref{out2}) follows from the relation
\begin{equation}\label{normpsi}
\sum_{k=0}^{N-1}\tilde g(k)\tilde g(q-k) = \frac1N\sum_{m=0}^{N-1}g^2(m)e^{-i2\pi qm/N} = \tilde g_1(q).
\end{equation}

Eq.(\ref{out2}) represents a Schmidt decomposition of the two-mode entangled state at the output of the beam-splitter. It allows us to compute the entanglement of formation:
\begin{equation}\label{E}
E=-\sum_{k=0}^{N-1}\frac{\tilde g(k)\tilde g(q-k)}{\tilde g_1(q)} \log_2\left\{\frac{\tilde g(k)\tilde g(q-k)}{\tilde g_1(q)}\right\},
\end{equation}
and provides us with information on the Schmidt basis for both modes. It is remarkable that the Schmidt basis for each mode given by a set of RICS, Eq.(\ref{alphaNm}), does not depend on $q$, but the correspondence of states of two modes does.

\subsection{Entanglement of a two-mode RICS as a function of the coherent amplitude and the number of components}
Entanglement determined by Eq.(\ref{E}) is in general a function of three variables: coherent amplitude $|\alpha_0|$, number of components $N$, and the ``rotational quantum number'' $q$. It can be easily computed with the help of Eqs.(\ref{g}) and (\ref{DFT}). In this subsection we will analyze its behavior keeping in mind that the bipartite entangled state under consideration can be created by splitting its corresponding single-mode ``in-state'' $|\mathrm{c}_q\rangle$ on a 50:50 beam splitter.

\begin{figure}[ht]
 \includegraphics[width=\columnwidth]{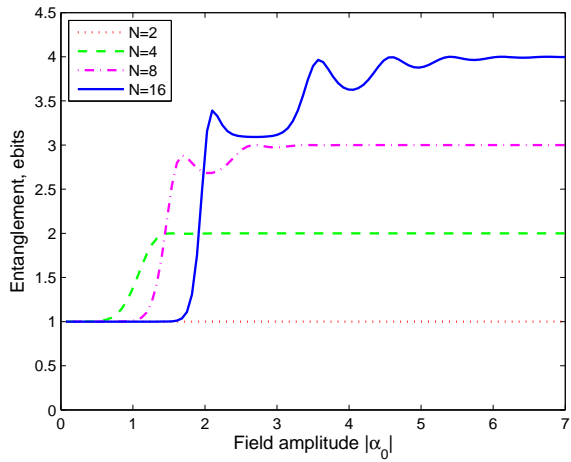}
\caption{\label{Ea-q0} Entanglement of formation for a two-mode RICS as function of the coherent amplitude for various numbers of components and $q=1$. The asymptotic value in all cases is $\log_2N$.}%
\end{figure}

In Fig.\ref{Ea-q0} we show the dependence of entanglement on the field amplitude for the case of $q=1$.

We see in Fig.\ref{Ea-q0} that for relatively small amplitude the entanglement is close to 1 ebit for any number of components, which is expected because the ``in-state'' $|\mathrm{c}_0\rangle$ is close, according to Eq.(\ref{lim}), to the single-photon state, which results in the state $2^{-1/2}(|1\rangle_A|0\rangle_B + |0\rangle_A|1\rangle_B)$ at the output of the beam splitter, having exactly 1 ebit of entanglement. For sufficiently large amplitude the components become almost orthogonal and the entanglement tends to $\log_2N$. At intermediate values of the amplitude an interference pattern is observed for a sufficiently large number of components. In the case of two components entanglement is independent of the amplitude, which is a well-known fact for the beam-split odd coherent state \cite{Sanders12}.

\begin{figure}[h]
\centerline{\includegraphics[width=1.0\columnwidth]{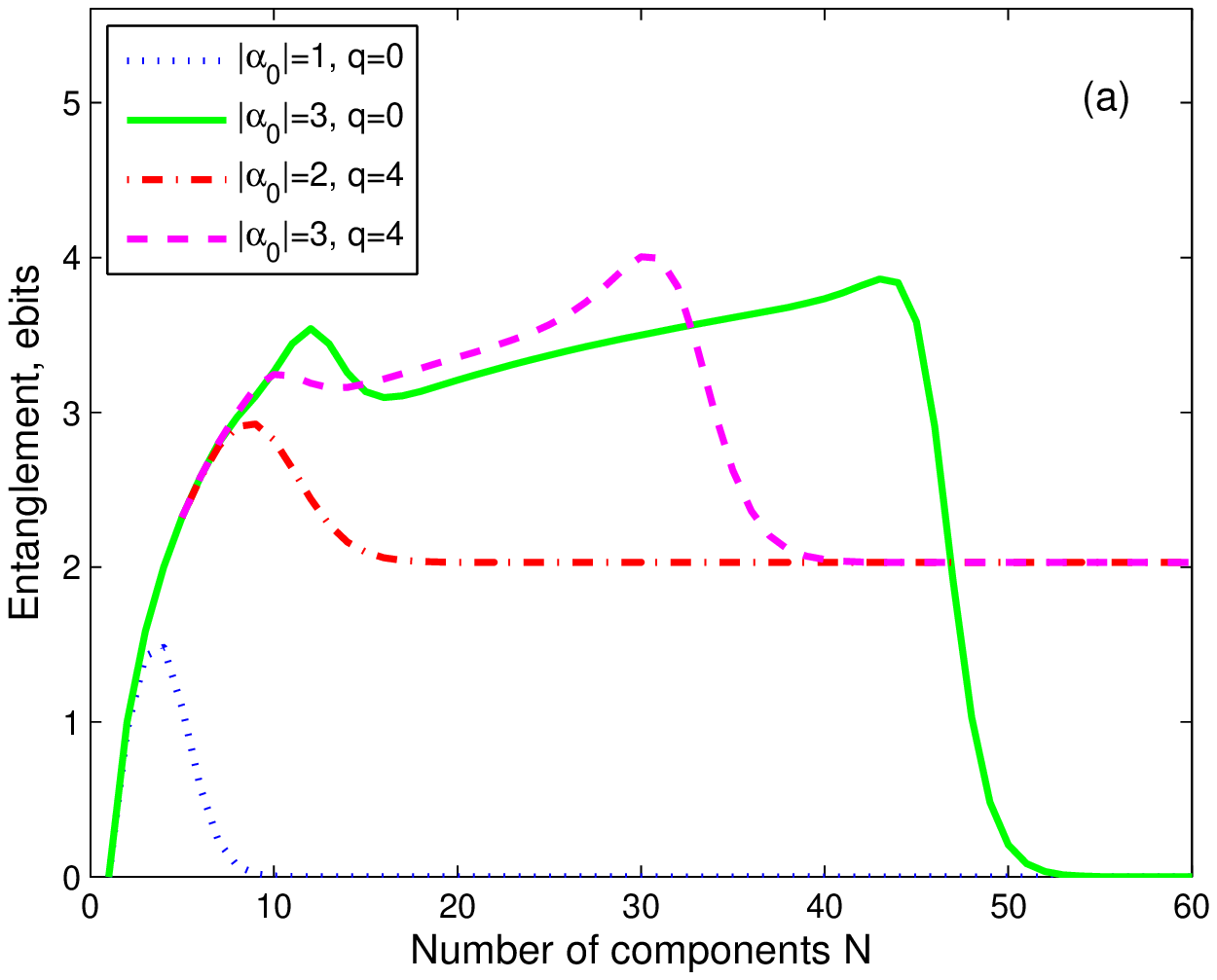}}
\centerline{\includegraphics[width=1.0\columnwidth]{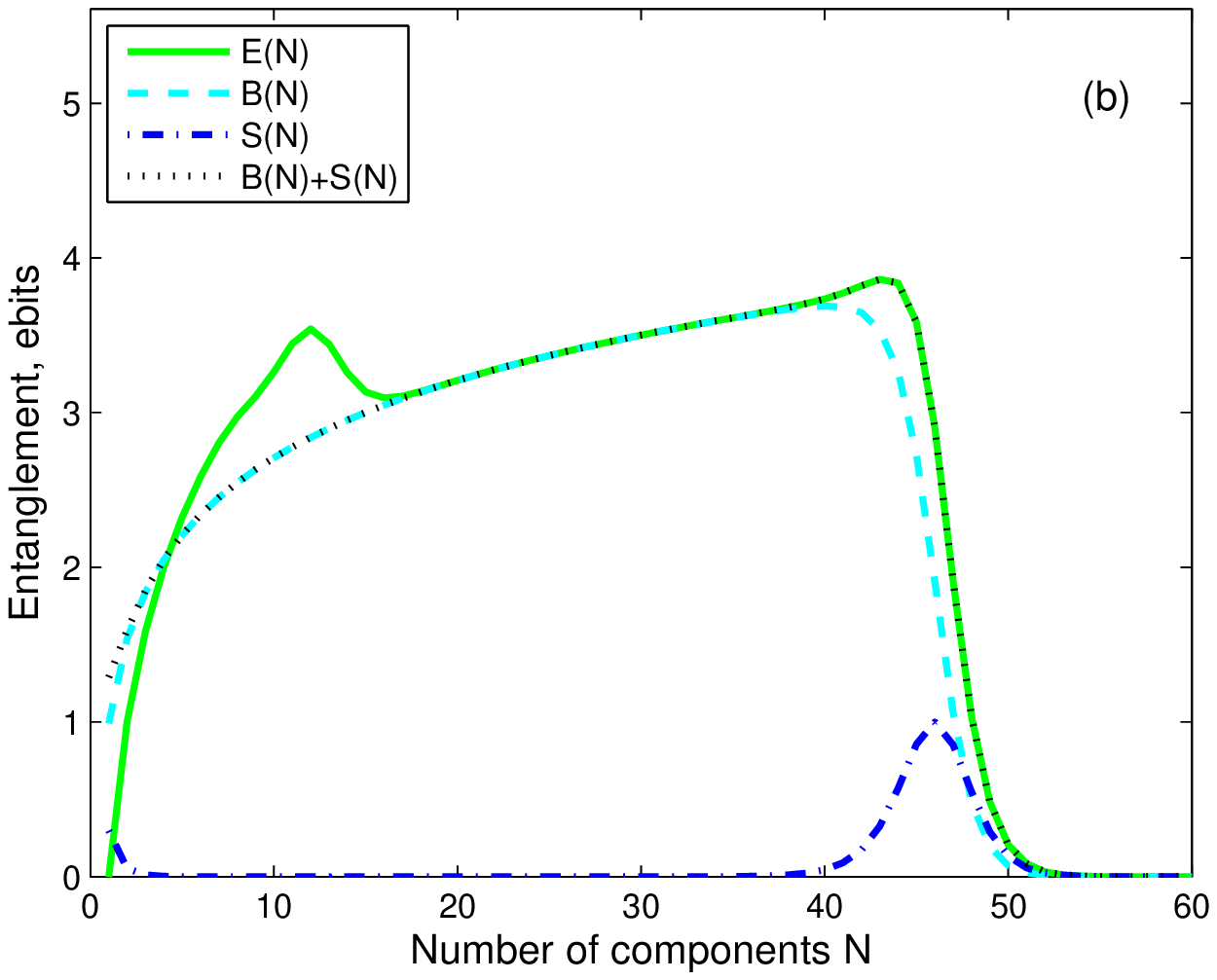}}
\caption{\label{EN-q0} Entanglement of formation for a two-mode RICS as a function of the number of components (a) for various values of $|\alpha_0|$ and $q$, and (b) for the case $|\alpha_0|=3$, $q=4$, represented as a sum of entropies.}%
\end{figure}

Fig.\ref{EN-q0}a shows the dependence of entanglement on the number of components. Entanglement increases logarithmically with growing $N$ until the states on the circle start to overlap, which happens approximately for $N\approx N_1=\pi |\alpha_0|$, i.e. when $2\sigma$ areas of the neighbor states on the circle start to overlap. For sufficiently large $N$ the entanglement is a constant, whose value becomes clear after noting that the ``in-state'' $|\mathrm{c}_q\rangle$ tends to the Fock state $|q\rangle$ and a beam-split Fock state is a binomial state of two modes:
\begin{equation}\label{transFock}
|q\rangle_A|0\rangle_B \rightarrow \sum_{m=0}^q\sqrt\frac{q!\,2^{-q}}{m!(q-m)!}|m\rangle_A|q-m\rangle_B,
\end{equation}
So, the entanglement is given by the entropy of binomial distribution $P(k)=\frac{q!}{k!(q-k)!}2^{-q}$, which for $q=4$ is close to 2.047.

An interesting feature is the abrupt drop of entanglement just before it takes its constant asymptotic value. Such behaviour can be explained by considering  the two first terms in Eq.(\ref{cq}) for the ``in-state'', for example for $q=0$:
\begin{equation}\label{cx}
|\mathrm{c}_0\rangle\approx \frac1{\sqrt{1+X}}\left(|0\rangle+\sqrt{X}|N\rangle\right),
\end{equation}
where $X=(2|\alpha_0|^2)^{N}/N!$ is a coefficient experiencing a fast decay in the region $N > 2|\alpha_0|^2$. After the beam-splitter such a state results in a superposition of the two-mode vacuum and a binomial state, which for $N\gg 1$ are almost biorthogonal \cite{Linden06}. It can be shown by applying Eq.(\ref{transFock}) to Eq.(\ref{cx}) and a direct calculation, that the entanglement after the beam-splitter can be written approximately as $E(N)=B(N)+S(N)$, where
\begin{equation}\label{BN}
B(N)=\frac{X}{1+X}\frac12\log_2\left(\frac{\pi e N}2\right),
\end{equation}
is the entropy of the binomial distribution times its weight, while
\begin{equation}\label{SN}
S(N)=-\frac{1}{1+X}\log_2\frac{1}{1+X}-\frac{X}{1+X}\log_2\frac{X}{1+X},
\end{equation}
is the Shannon entropy of the distribution of weights in Eq.(\ref{cx}). In Fig.\ref{EN-q0}b we show both entropies and their sum, which describes well the behaviour of the entanglement in the region $N>2|\alpha_0|^2=18$.

We see that the growth of entanglement before the last local maximum is given by the growing entropy of the binomial distribution. The maximum itself occurs when both terms in Eq.(\ref{cx}) become comparable and $S(N)$ becomes significant. The value of $N$ such that $X=1$ can be taken as the limiting value, after which $E(N)$ is almost constant. For the considered example of $q=0$ this value is approximately $N_2=2e|\alpha_0|^2$. Above this value the vacuum in Eq.(\ref{cx}) becomes dominating and the entanglement of the corresponding two-mode state is zero. The sum of entropies does not explain the first local maximum, which requires the additional term $|2N\rangle$ in Eq.(\ref{cx}).

\section{RICS as a basis}
\subsection{Decomposition of a circular state}
The family of RICS, Eq.(\ref{alphaNm}), for any fixed $N$ and $\alpha_0$ form an orthonormal basis in the $N$-dimensional subspace $\mathcal{S}_N$ of the state space of quantum harmonic oscillator, and the states from  $\mathcal{S}_N$ can be decomposed in this basis. In particular, for a coherent state $|\alpha_m\rangle$ we obtain from Eq.(\ref{alphaNm})
\begin{equation}\label{coherent}
|\alpha_m\rangle = \sum_{q=0}^{N-1}\sqrt{\tilde g(q)}e^{-i2\pi qm/N}|c_q\rangle,
\end{equation}

Any $N$-component circular state, Eq.(\ref{circ}), with arbitrary coefficients $c_m$ can be represented as a superposition of RICS:
\begin{equation}\label{superp}
|\psi_c\rangle = N\sum_{q=0}^{N-1}\sqrt{\tilde g(q)}\tilde c_q|c_q\rangle,
\end{equation}
with a normalization condition
\begin{equation}\label{norm}
\sum_{q=0}^{N-1}\tilde g(q)|\tilde c_q|^2=\frac1{N^2}.
\end{equation}

\subsection{Quadratic phase}

A RICS, defined by Eq.(\ref{alphaNm}) is an equally weighted circular state with a linear relative phase dependence. Another interesting case is an equally weighted superposition with a quadratic phase. Let us consider a circular state Eq.(\ref{circ}) with the following coefficients \cite{Tara93}
\begin{equation}\label{kerr}
\tilde c_k = \frac1N e^{-i\pi k(k-p)/N},
\end{equation}
where $p=(N\mod2)$ is the parity of $N$. Such coefficients characterize a family of ``Kerr states'', being the subject of a vast literature since mid-80s \cite{Yurke86,Buzek92,Horoshko97,Tanas91,Horoshko98,vanEnk03}. The states of this family are generated by a third-order (Kerr) nonlinearity in the microwave spectral domain for mesoscopic values of the field amplitude, up to $|\alpha_0|=10$ \cite{Kirchmair15}. It is worth noting that in the direct domain the coefficients $\{c_n\}$ are also equally weighted and have a quadratic phase depedence \cite{vanEnk03}.

When a Kerr state is split on a beam-splitter, or considered in a transformed modal basis, it produces a two-mode circular state. The partial density matrix of one mode in the basis of RICS is given by substituting Eq.(\ref{kerr}) into Eq.(\ref{Ra}):
\begin{equation}\label{Rkerr}
\hat R'_{mn} = \sqrt{\tilde g(m)\tilde g(n)} g(n-m) e^{i\pi(n(n-p)-m(m-p))/N},
\end{equation}
and the entanglement of the two modes can be found numerically as the Shannon entropy of the eigenvalues of this matrix, which has a very simple form.

\subsection{Entanglement of a general two-mode circular state}
The case of a general two-mode circular state can be treated numerically by finding the eigenvalues $\{\lambda_k\}$ of the $N\times N$ partial density matrix of one mode, Eq.(\ref{Ra}).

Alternatively, the eigenvalue problem for the partial density matrix after the beam-splitter can be solved in a truncated Fock basis, which is a traditional approach to the problem. When the ``in-state'' is represented as a sum of Fock states from $|0\rangle$ to $|K-1\rangle$, where $K>2|\alpha_0|^2+|\alpha_0|$, the partial density operator of the corresponding two-mode state is represented by an $K\times K$ matrix. However, in many interesting cases $K\gg N$, and in the truncated Fock basis the numerical solution gives approximate eigenvalues $\{\lambda'_k\}$ of an approximate matrix, while in the RICS basis it gives approximate eigenvalues of the \emph{exact} matrix Eq.(\ref{Ra}), providing a higher degree of precision.

Numerical treatment of the eigenvalue problem for the partial density operator is also possible in the non-orthogonal basis of coherent states $\{|\alpha_m\rangle, m=0,1,...,N-1\}$ \cite{Janszky03}, where the eigenvalues of a $N\times N$ matrix are to be found. However, RICS have an advantage of being an orthogonal basis, which is much more convenient for the practical calculations.

It is known that a circular state with properly chosen parameters $\{N,\alpha_0\}$ can approximate many (but not all) states of the single-mode field with arbitrary precision \cite{Szabo96}. If such a state is split on a beam-splitter, the obtained two-mode state can be approximated by a two-mode circular state. It is natural to use the RICS basis for performing the numerical calculation of the entanglement of this two-mode state.

\section{Discussion}

In this section we will discuss two questions: the optimal entanglement of a two-mode circular state for given resources and the quantification of non-classicality of the in-state by the entanglement of the corresponding out-state.

The first question we address is ``How much entanglement can be produced by splitting on a beam splitter a general circular state, defined by Eq.(\ref{circ}) with a fixed $|\alpha_0|$?''. Thus, we consider the coherent amplitude $|\alpha_0|$ as a resource and would like to know how this resource can be optimally used for producing an optimal coherent superposition. First of all we note that the entanglement of the state Eq.(\ref{Psid}) is not greater than $\log_2N$, because $N$ is the rank of the partial density operator, given by Eq.(\ref{rhod}).

\begin{figure}[h]
\centerline{\includegraphics[width=1.0\columnwidth]{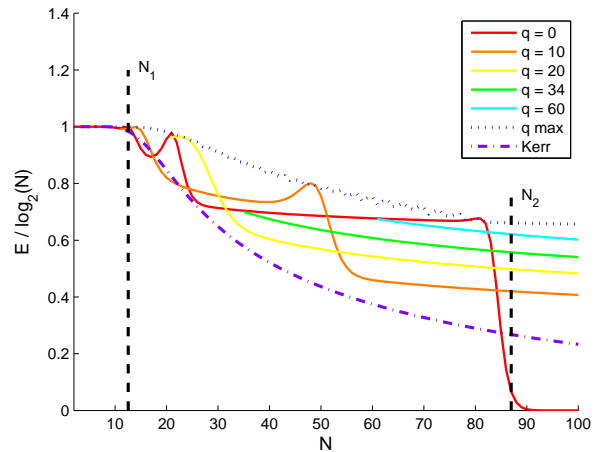}}
\caption{\label{EN-a4} Entanglement of formation for a two-mode RICS, $|\alpha_0|=4$ and for the Kerr state with the same amplitude. The dependence has three regions: (i) a logarithmic growth up to $N_1=\pi|\alpha_0|$ while the components are almost orthogonal; (ii) oscillations when the components start to overlap and interfere; (iii) asymptotically constant value after $N_2=2e|\alpha_0|^2$. q~max corresponds to maximal entanglement for given $N$.}
\end{figure}

Considerable amount of entanglement can be obtained by splitting a RICS. Let us analyse how the entanglement of a two-mode RICS depends on the number $N$ of components. As has been already found in Sec.~III, this dependence has three regions (see also Fig.~\ref{EN-a4}): (i) a logarithmic growth up to $N_1=\pi|\alpha_0|$ while the components are almost orthogonal; (ii) oscillations when the components start to overlap and interfere; (iii) asymptotically constant value after $N_2=2e|\alpha_0|^2$, when entanglement is given by the entropy of the binomial distribution with parameter $q$, which is approximately
\begin{equation}\label{Ebin}
E_{bin}(q)=\frac12\log_2(\pi e q/2)
\end{equation}
for $q\gg 1$. The first region corresponds to the the so-called ``Schr\"odinger cat state'' for the ``in-state'', while the third region corresponds to approaching the Fock state $|\psi_c^{in}\rangle_{A}=|q\rangle_A$.

In general, the maximal entanglement for a RICS state is limited by the following bounds:
\begin{equation}\label{ineq}
\frac12\log_2(N) < \max_q E(|\alpha_0|,N,q) \le \log_2(N),
\end{equation}
where the lower bound in the first and second regions has been verified numerically for experimentally interesting values $0<|\alpha_0|\le 4$, while in the third region it is always surpassed by a RICS with $q=N-1$, as can be seen from Eq.~(\ref{Ebin}).

As can be seen from Fig.~\ref{EN-a4}, the lower bound in Eq.(\ref{ineq}) is not optimal for $N<N_2$, where $E(|\alpha_0|, N, q_\mathrm{max})$ is substantially larger than $\frac12\log_2N$. An interesting question is what is the maximal entanglement of formation $E_{\mathrm{max}}^{\mathrm{CS}}(N)$, at fixed $N$, for all  circular states with given fixed $\alpha_0$, but arbitrary $c_m$. Clearly, $E_{\mathrm{max}}^{\mathrm{CS}}(N)\leq \log_2N$. Our analysis above of the entanglement of formation of the RICS shows it is at least as large as $\frac12 \log_2N$. It is not clear whether other choices of the $c_m$, possible with non-constant $|c_m|$, would yield a higher entropy of formation.

One interesting consequence of Eq.~(\ref{ineq}) is that the limiting bounds of entanglement of a RICS state are determined not by the coherent amplitude $|\alpha_0|$ but by the number of components $N$. This behavior is drastically different as compared with other circular states such as, for example, Kerr states~\cite{vanEnk03}. For each $\alpha_0$ and $N$ there is only one Kerr state, and the maximal over $N$ entanglement is limited by the value of $|\alpha_0|$ and is approximately equal to $\log_2(\pi|\alpha_0|)$ (see also Fig.~\ref{EN-a4}). The explanation of this difference can be given on the basis of consideration of the mean energy or the mean photon number.

Kerr states are generated deterministically from a coherent state with the amplitude $\alpha_0$, and have the same mean photon number $|\alpha_0|^2$. RICS states can be produced from a coherent state by two different ways: probabilistically, ~i.~e. via random projection of coherent state $\alpha_0$ on a RICS state, or deterministically like Kerr states. Let us consider first a probabilistic method \cite{Raimond97,Domokos94}. In this case the mean photon number of a RICS state is  given by Eq.~(\ref{alphaNm}) as
\begin{equation}\label{mean}
\bar n(q) \equiv \langle\mathrm{c}_q|a^\dagger a|\mathrm{c}_q\rangle = |\alpha_0|^2\frac{\tilde g(q-1)}{\tilde g(q)}\approx q,
\end{equation}
where the approximate equality holds for $N\gg N_2$. This result demonstrates that for $q\gg |\alpha_0|$ one can obtain a RICS state with the mean photon number much larger than the average number of photons $|\alpha_0|^2$ in the corresponding coherent state. However, the probability of obtaining such an outcome is very low. Indeed, the probability of obtaining a RICS from a coherent state by a projection is given by Eq.~(\ref{coherent}) as $p(q)=|\langle \alpha_m|\mathrm{c}_q\rangle|^2=\tilde g(q)$ and is rapidly decreasing for high values of $q$. Calculating the average photon number over many projections we obtain $\sum p(q)\bar n(q) = |\alpha_0|^2$. Thus, a random projection generates a RICS with a random photon number, which sometimes can be very high. For the same state entanglement of the ``out-state'' is also very high. However the probability of obtaining such high entanglement is very low.

In the case of deterministic generation of RICS, Eq.~(\ref{mean}) for the mean number of photons remains valid. However, since in this case a RICS with $\bar n(q) \approx q$, $0\le q\le N-1$, is created deterministically, the interpretation of the result is different from the probabilistic case. Indeed, when $q > |\alpha_0|^2$ one needs some additional external energy in order to create such a state. Therefore, the initial coherent amplitude $|\alpha_0|$ cannot be considered as the only resource for creating entanglement with the value given by Eq.~(\ref{ineq}).

The second question of our interest is the quantification of non-classicality of a single-mode state, i.e. introduction of a measure, which is zero for classical states (coherent states and their mixtures) and is positive for non-classical states. For a pure ``in-state'', it is demonstrated in Ref.~\cite{Asboth05} that the entanglement $E$ of the ``out-state'' can be an appropriate measure of nonclassicality. Therefore, we can interpret our Eq.~(\ref{E}) as a measure of nonclassicality of a RICS and compare it with alternative nonclassicality measures. For example, one of such alternative measures is given simply by a number of coherent states necessary for the composition of a single-mode quantum state \cite{Vogel14}. For a RICS this number is equal to $N$. We have observed that, when applied to RICS states, these two measures can predict two qualitatively different results. Precisely, for $q=0$, $\alpha_0$ fixed and $N$ becoming very large, the measure of nonclassicality from Ref.~\cite{Vogel14} grows respectively as $N$, thus predicting an increase of nonclassicality. On the other hand, since the entanglement of formation $E$ of the ``out-state'' is tending to zero for this sequence, the corresponding entropic measure of nonclassicality predicts approaching a classical state. In our opinion, this prediction has a much higher practical significance, because in this limit the sequence of RICS tends to the classical state of vacuum.

\section{Conclusions}
In the present work we have approached the problem of calculating the entanglement between two modes of a two-mode circular state of light. We have found, that for a subclass of two-mode circular states, which we call two-mode RICS, an analytical solution for the Schmidt decomposition exists, and therefore a rather simple analytical expression gives the exact value of the entanglement, Eq.(\ref{E}). On the basis of this exact expression we have investigated the behaviour of the entanglement of a two-mode RICS and found its main features. We have shown that a set of single-mode RICS form a basis in the space of circular states, and this basis is advantageous for calculating entanglement of any two-mode circular state. We have shown also that the maximal attainable entanglement of an arbitrary two-mode circular state on a circle of given radius grows logarithmically with the number of components.

\acknowledgments

This work was supported in part by the Labex CEMPI (ANR-11-LABX-0007-01) and by the Nord-Pas de Calais Regional Council and FEDER through the Contrat de Projets \'Etat-R\'egion (CPER), in part by  the European Union's Horizon 2020 research and innovation programme under grant agreement No 665148 and in part by Belarusian Republican Foundation for Fundamental Research. D.H thanks the Laboratoire PhLAM, the CEMPI and the FEDER for their hospitality and for funding his stay at the University Lille 1, which made this work possible. The authors thank D. Spehner for stimulating discussions on the subject matter of this paper.


\begin{thebibliography}{00}
\bibitem{ScullyZubairy97} M. O. Scully and M. S. Zubairy, Quantum Optics (Cambridge University Press, 1997).
\bibitem{Dodonov74}V.V.Dodonov, I. A. Malkin, and V. I. Man'ko, Physica (Utrecht) 72, 597 (1974).
\bibitem{Yurke86}B. Yurke and B. Stoler, Phys. Rev. Lett. 57, 13 (1986).
\bibitem{Buzek92}V. Buzek, A. Vidiella-Barranco, and P. L. Knight, Phys. Rev. A 45, 6570 (1992).
\bibitem{Horoshko97}D.B.Horoshko and S.Ya.Kilin, Phys. Rev. Lett. 78, 840 (1997).
\bibitem{Raimond97}J. M. Raimond, M. Brune, and S. Haroche, Phys. Rev. Lett. 79, 1964 (1997).
\bibitem{Kirchmair15}G. Kirchmair et al., Nature 495, 206 (2013).
\bibitem{Takahashi08}H. Takahashi et al., Phys. Rev. Lett. 101, 233605 (2008).
\bibitem{Ourjoumtsev09}A. Ourjoumtsev, F. Ferreyrol, R. Tualle-Brouri, P. Grangier, Nature Physics, 5, 189 (2009).
%\bibitem{Prasad87} S. Prasad, M. O. Scully, W. Martienssen, Opt. Comm. 62, 139 (1987).
\bibitem{Sanders92}B. C. Sanders, Phys. Rev. A 45, 6811 (1992).
\bibitem{vanEnk03}S. J. van Enk, Phys. Rev. Lett. 91, 017902 (2003).
\bibitem{Kilin11}S. Ya. Kilin and A. B. Mikhalychev, Phys. Rev. A 83, 052303 (2011).
\bibitem{Sanders12}B. C. Sanders, J. Phys. A: Math. Theor. 45, 244002 (2012).
\bibitem{Bennett96} C. H. Bennett, D. P. DiVincenzo, J. Smolin, and W. K. Wootters, Phys. Rev. A 54, 3824 (1996).
\bibitem{Law00} C. K. Law, I.A. Walmsley, and J. H. Eberly,
Phys. Rev. Lett. \textbf{84}, (2000) 5304.
\bibitem{Jose00}W. D. Jose and S. S. Mizrahi, J. Opt. B: Quantum Semiclass. Opt. 2, 306 (2000).
\bibitem{Wang10}Y.-Y. Wang, Z.-J. Liu, Q.-H. Liao, and S.-T. Liu, Chin. Phys. B 19, 054204 (2010).
\bibitem{Horoshko98}D.B.Horoshko and S.Ya.Kilin, Opt. Expr. 2, 347 (1998).
\bibitem{Zurek01} W. H. Zurek, Nature 412, 712 (2001).
\bibitem{Birulia68}Z. Bialynicka-Birulia, Phys. Rev. 17, 1207 (1968).
%\bibitem{Miranowicz90}A. Miranowicz, R. Tanas, and S. Kielich, Quantum Opt. 2, 253 (1990).
\bibitem{Tanas91}R. Tanas, T. Gantsog, A. Miranowicz, and S. Kielich, J. Opt. Soc. Am. B 8, 1576 (1991).
\bibitem{Tara93}K. Tara, G. S. Agarwal, and S. Chaturvedi, Phys. Rev. A 47, 5024 (1993).
\bibitem{Janszky93}J. Janszky, P. Domokos, and P. Adam, Phys. Rev. A 48, 2213 (1993).
\bibitem{Janszky95}J. Janszky, P. Domokos, S. Szabo, and P. Adam, Phys. Rev. A 51, 4191 (1995).
\bibitem{Domokos94}P. Domokos, J. Janszky, and P. Adam, Phys. Rev. A 50, 3340 (1994).
\bibitem{Asboth05}J. K. Asboth, J. Calsamiglia, and H. Ritsch, Phys. Rev. Lett. 94, 173602 (2005).
\bibitem{Lowdin50}P.-O.~L\"owdin, J. Chem. Phys. 18, 365 (1950).
\bibitem{Mayer}I.~Mayer, Int. J. Quantum Chem. 90, 63 (2002).
\bibitem{Janszky03}J. Janszky et al., Fortschritte der Physik, 51, 157 (2003).
\bibitem{Szabo96}S. Szabo, P. Adam, J. Janszky, and P. Domokos, Phys. Rev. A 53, 2698 (1996).
\bibitem{Linden06}N. Linden, S. Popescu, and J. A. Smolin, Phys. Rev. Lett. 97, 100502 (2006).
\bibitem{Vogel14}W. Vogel and J. Sperling, Phys. Rev. A 89, 052302 (2014).

\end{thebibliography}
\end{document}